# On the gravitodynamics of moving bodies

**Anderson W. Mol**

Universidade Estadual de Santa Cruz, DCET–CPqCTR, 45662−900 Ilhéus, Brazil. [†]

Universidade do Porto, Centro de Física do Porto, 4169−007 Porto, Portugal.

Universidade de Aveiro, Departamento de Física, i3N, 3810−193 Aveiro, Portugal.

**Abstract**

In the present work we propose a generalization of Newton's gravitational theory from the original works of Heaviside and Sciama, that takes into account both approaches, and accomplishes the same result in a simpler way than the standard cosmological approach. The established formulation describes the local gravitational field related to the observables and effectively implements the Mach's principle in a quantitative form that retakes Dirac's large number hypothesis. As a consequence of the equivalence principle and the application of this formulation to the observable universe, we obtain, as an immediate result, a value of $\Omega = 2$. We construct a dynamic model for a galaxy without dark matter, which fits well with recent observational data, in terms of a variable effective inertial mass that reflects the present dynamic state of the universe and that replicates from first principles, the phenomenology proposed in MOND. The remarkable aspect of these results is the connection of the effect dubbed dark matter with the dark energy field, which makes it possible for us to interpret it as longitudinal gravitational waves.

## 1   Introduction

Newton's laws of motion and the theory of universal gravitation constitute, even today, the epistemological base of our understanding of physical science. Because of the spectacular success of Newtonian Mechanics, for a long period of time, it was generally felt that the laws and the *conceptual basis* in which it rests needed not to be reinterpreted or submitted to any critical review. This status perdured until the beginning of the last century, when it was widely recognized that Newtonian Mechanics needed to be replaced by quantum mechanics and relativistic mechanics, depending on whether the size of objects considered are too small or the speeds involved are comparable to that of light and/or the intensity of the gravitational field is too strong. But in the absence of these conditions, i.e. in the domain of classical physics, Newton's laws are believed to represent, to a high degree of accuracy, the limits of these theories. One of them is in the galactic scale where the speeds of the objects, i.e. stars and interstellar clouds of gases and dust, are non-relativistic and the gravitational fields are extremely weak. Under these conditions we would expect that classical dynamics and gravitation were extremely successful, but they aren't! One such problem that arises is represented by the galaxy's flat rotation curve. This problem could be well understood if we consider a large amount of matter, e.g. the galaxy bulge and a star revolving around it in a quasi circular orbit. From the Newtonian point of view we know that if we increase such amount of matter the rotation speed would increase as well, to compensate for it. On the other hand, if we consider stars more distant from the centre we would expect to observe a decrease in the orbital speed in accordance with what is known as the Keplerian regime, which is well verified in the solar system. However, for the most part of the observed galaxies, this characteristic velocities profile is surpassed by far, beyond what could be possibly explained by considering only the total amount of observed baryonic matter in all its known form. This situation is so remarkable that the amount of missing mass needed to be taken into account for the observed dynamical effects is up to one order of magnitude. To preserve the classical dynamics a new kind of matter is supposed to exist in an unknown form that neither emits nor reflects light. This constitutes the dark matter problem. Currently, the mainstream in physics endorses this approach, i.e. that this is the right answer to the riddle. As commented by L. Smolin ([1], pg. 15):

> 'The dark-matter hypothesis is preferred mostly because *the only other possibility* — that we are wrong about Newton's laws, and by extension *general relativity* — is too scary to contemplate.'

In the present work all the words and phrases in italics are our own highlights and do not necessarily reflect the original intentions of the quoted authors.

If we observe the universe beyond the cluster and supercluster scale, i.e. at a very large scale corresponding to billions of light-years, the problem becomes more puzzling. The results of the observations indicate that the expansion of the universe is speeding up, instead of slowing down owing to the mutual gravitational attraction due to the observed matter. The equations of the general theory of relativity (GTR) are not satisfied even when the estimated amount of dark matter is added in. In fact, as a result, it should be doing the opposite — decelerating. Perhaps when one gets to a scale comparable to the size of the universe, GTR is simply no longer applicable. This indicates that there is much more to the universe than we understand at present [2]. The leading interpretation is that the universe is filled by something dubbed dark energy that antigravitates. Whereas the possibility for gravitational repulsion does not exist in Newtonian gravity, it does exist in general relativity. The equivalence between matter and energy suggests that a new kind of matter/energy that actuates as an energy density fluid with a sufficiently negative-pressure can be a source of a repulsive gravitational field. It has been realized that some of the quantum fields that arise in elementary-particle theory allow for fluids with negative pressure that will cause a repulsive gravity. The dark energy would be, thus, simply the effect of a negative-pressure fluid that is postulated to account for the present cosmic acceleration. The immediate candidate for dark energy is Einstein's cosmological constant $\Lambda$, which designates a perfectly uniform fluid with negative pressure that is associated with the lowest energy vacuum state of the universe. However, the observationally required value of the cosmological constant is $10^{120}$ times smaller than the theoretical expectation. This constitutes the dark energy problem.

Recent measurements from the Wilkinson Microwave Anisotropy Probe (WMAP) spacecraft, as currently interpreted, reveal a universe consisting mostly of the unknown [3]. In terms of their contribution to the mean energy density, the contents of the universe are approximately 72% dark energy, 24% dark matter and 4% ordinary baryonic matter, with smaller contributions from photons and neutrinos. So an amount on the order of 96% of the whole universe is at present absolutely unknown. These numbers could be interpreted too from another perspective, i.e. as an exact measure of our lack of knowledge about the true nature of gravitation.

After all is too hard to admit that, past more than three hundred years of success in gravitational physics, we presently face the prospect that we don't know yet what gravity actually is!

## 2 Mach's principle

The question of whether space is an independent entity with its own reality expression or only a mere subjective perception, being nothing more than what is specified: the distance between the experiential bodies, has been, since the ancient Greeks, a long tradition in the western philosophical reasoning about the true nature of space and time, and is probably rooted in our common and immediate perception about the distinction between the objects and the empty space amongst them. From these historical roots two distinct views about the nature of space and time emerge, considering them either as absolute or relative. The absolute view identifies the space as a container holding all material objects in which bodies can move, but which exists independently of its content, while the relative view considers space merely as a conceptual abstraction of the storage of the individual bodies that consequently looses its meaning without them [4].

The origin of this divergence in our modern scientific thought about the nature of space and time started 23 years after Newton published in 1687 his theory of inertia in *Principia*, when it was strongly attacked on philosophical bases, by the famous philosopher G. Berkeley [5]. Berkeley pointed that motion had meaning only when referenced to nearby objects and that there could not be such a *metaphysical absolute space* as proposed by Newton. Berkeley supported his point of view reanalyzing an experiment performed by Newton, a suspended rotating bucket.

This experiment consists of a bucket of water suspended by a rope, which is twisted so that upon release the bucket rapidly acquires rotation. This motion is soon communicated to the water that subsequently rotates forming a concave surface. Then the bucket is instantaneously stopped and held motionless, however the water continues to rotate for a while, keeping its concave surface. Progressively it comes back to rest and its surface becomes gradually flat again.

Newton performed this experiment in an attempt to resolve a basic difficulty in his second law of motion. This law states that the acceleration experienced by a body is equal to the force acting on it divided by its mass, or expressing it in standard form, that the force is equal to the mass times the acceleration. The trouble that follows immediately is: How shall we measure acceleration and with respect to which reference? To the earth itself? To the moon or to the sun? Despite the fact that both are accelerated with respect to us and to each other. Thus, to avoid the previous difficulties, Newton interpreted his experiment on the basis of the fact that the relative motion of the water and the bucket apparently did not affect the surface of the water, and so he postulated that there was such a thing as absolute space and that his second law of motion applies *only* to *absolute motion*. This leaves us with the paradox that acceleration has an intrinsic relative nature or a *relational* feature as exposed by A. K. T. Assis [6], since one can actually only observe relative motions.

The critic of E. Mach [7] does not differ in its essence from that of Berkeley's, and its main merit besides being more physically detailed than Berkeley's, resides on the fact that it was important in the moment that Newton's authority was unquestionable and started a process of rediscussion of the foundations of mechanics that leaded to the revolution in physics in the forthcoming years. Mach's criticism to Newton's laws in what concerns the inertial forces is summarized in the following quotation ([8], pg. 330):

> 'Obviously it does not matter if we think of the earth as turning round on its axis, or at rest while the fixed stars revolve round it. Geometrically these are exactly the same case of a relative rotation of the earth and the fixed stars with respect to one another. But if we think of the earth at rest and the fixed stars revolving round it, there is no flattering of the earth, no Foucault's experiment and so on—at least according to our usual conception of the law of inertia. Now one can solve the difficulty in two ways. Either all motion is absolute, or our law of inertia is wrongly expressed. I prefer the second way. *The law of inertia must be so conceived that exactly the same thing results from the second supposition as from the first*. By this it will be evident that in its expression, regard must be paid to the *masses of the universe*… All bodies, each with its share, are of importance for the law of inertia.'

Then, according to Mach, *there would not be any preferential frame of reference*, i.e. the laws of dynamics would be the same for inertial and non-inertial frames showing the same results. If Mach's concepts were right, and Newton's calculations were also correct, we would have to propose a re-formulation of the gravitation that effectively encompasses both aspects, qualitative and quantitative, to take into account the effects of distant masses and yet more importantly, to explain conceptually how that could be so, if we have to consider a finite velocity for the propagation of gravitational interaction and its consistency with the established relativistic principles.

Concerning to the previous arguments and the present knowledge of the physical theories we are invited to reflect on a simple example to illustrate how Mach's principle works and its compatibility with the well-known physical principles. To proceed with such analysis let us consider the solar system, with its planets turning around its centre under the influence of the sun's gravitational field, and considering that the earth rotates around it at a distance that a beam of light spends approximately 8 min 20'' to cross; to be rigorous this is the minimum time interval required by a physical sign to traverse such a distance at a maximum possible physical speed $c$. Let us perform a gedanken experiment in which we own a remote control that enables us to switch-off the sun in *every respect*: If we go on with, and point the control toward the sun to completely turn it off, after this time interval the electromagnetic wave sign would have reached the sun, and again after the elapse of the same period of time the sun's light will have disappeared to us, explicitly the earth will be in darkness. What can we say about the sun's gravity? After the GTR we have good arguments to presume that the same will be true for gravity, i.e. after the same period of time and not before, the earth will be liberated of its orbit to follow in a straight line towards the outer space. The argument that justifies this reasoning is our knowledge that gravitation propagates as well through gravitational waves at the same maximum speed *c*, and for that reason it takes time to happen, it is not *instantaneous*! Therefore we would observe that the sunlight turns off at the same time that the sun's gravitational field is *locally extinguished*, freeing the planet of its bound trajectory. The remarkable fact is that throughout this time interval, the earth revolved around a centre of force where presumably the sun wasn't supposed to be any longer, and the immediate conclusion is that it would existed to us, as observers on earth, a *local gravitational field* that yet corresponded to the existence of the sun in the centre in which the earth was under its influence, while the changing information of this physical condition is not observed for us. If we lead the previous conclusion to the utmost extension, we can assume that this will be true for the whole *observable universe*, i.e. for the closest stars of our own galaxy to the proto-galaxies and quasars on the edge of the cosmic horizon, since each of them contributes with its share to this local field, determining our observed physical effects. Therefore this implies that *our empirical description of the physical reality is determined by what we observe here and now instead of what we anticipate it to be*. It doesn't matter which is the dynamical state of each material body in its own time; what is physically meaningful to us is what we observe locally from them in our present time. Under this perspective, the Mach's principle becomes understandable and fully compatible with the notion of absolute space and the moving bodies themselves, i.e. the space is completely filled by the corresponding *dynamic field* of each distant material body *from the past*, and we accelerate effectively with respect to this present local field originated from them ever since they become observable to us. The simultaneity of the physical events to us is not only mere images of the distant past actions but the proper physical reality that reaches us determining our own *objective reality* with its physical laws, properties and effects. From this point of view, almost paradoxical we are interacting *instantaneously* with the past, i.e. experiencing locally the influence of these distant material bodies through *space* and *time*; thus without violating the principle of causality.

A. Einstein gave an address [9] on May 5, 1920 at the University of Leiden. He chose as his theme Ether and the Theory of Relativity. He lectured in German, but we present below an English translation of two brief quotes that summarize his change of view about the true nature of space:

> 'If we consider the gravitational field and the electromagnetic field from the standpoint of the ether hypothesis, we find a remarkable difference between the two. There can be no space nor any part of space without gravitational potentials; for these confer upon space its metrical qualities, without which it cannot be imagined at all. *The existence of the gravitational field is inseparably bound up with the existence of space*. On the other hand a part of space may very well be imagined without an electromagnetic field; thus in contrast with the gravitational field, the electromagnetic field seems to be only secondarily linked to the ether, the formal nature of the electromagnetic field being as yet in no way determined by that of gravitational ether. From the present state of theory it looks as if the electromagnetic field, as opposed to the gravitational field, rests upon an entirely new formal motif, *as though nature might just as well have endowed the gravitational ether with fields of quite another type, for example, with fields of a scalar potential, instead of fields of the electromagnetic type*.
>
> ...Recapitulating, we may say that *according to the general theory of relativity space is endowed with physical qualities; in this sense, therefore, there exists an ether. According to the general theory of relativity space without ether is unthinkable*; for in such space there not only would be no propagation of light, but also no possibility of existence for standards of space and time (measuring-rods and clocks), nor therefore any space-time intervals in the physical sense. But this ether may not be thought of as endowed with the quality characteristic of ponderable media, as consisting of parts which may be tracked through time. The idea of motion may not be applied to it.'

Therefore we have to comment that the notion of space endowed with some physical quality is not new in physics, we must be aware in the light of new observational data that we need to look carefully and thoughtfully at this concept and do not reject it in advance by prejudice.

## 3  A brief review of the seminal papers and beyond

Our generalization of Newton's gravitational theory is based in *some extent* on the original works of O. Heaviside [10], and D. W. Sciama [11]. Although this important article of Heaviside was published in 1893, since then it appears to have been generally ignored (L. Brillouin [12], pp. 103-104 cites a reprint of this article), and his theory and results are practically unknown even today. The entire article has been recently reproduced, in modern notation, by O. Jefimenko ([13], pp.189-202), in appendix 8. As pointed out by Jefimenko, Heaviside's gravitational theory was based on equations practically identical to Maxwell's curl equations for electric and magnetic fields. These equations were universally believed to describe the phenomenon of electromagnetic induction. But at the time it was almost impossible to imagine that there could be anything similar in the domain of gravitation. There was nothing known in gravitation that could resemble, even remotely, electromagnetic induction. Another possibility of why the work of Heaviside did not call attention can be attributed to the fact that it was not fully developed, and soon it was overwhelmed by the GTR.

However we have to point out the importance of this work in what concerns the fact that for the first time the Newtonian theory of gravitation was conducted into the realm of time domain ([13], pg. 91), i.e. for a system that depends on the propagation time of the gravitational interaction, yielding results that were considered heretofore exclusive of the GTR.

In his article, Heaviside obtained the equation for a *velocity dependent gravitational field*, $\vec{g}_v$ as seen by an observer (Fig. 1), of a uniformly moving point mass $m$ in terms of its *retarded position* $\vec{r}$ respect to this observer taken as a reference. In modern notation this equation is:

$$\vec{g}_v = G \frac{m\left(1 - \frac{v^2}{c^2}\right)}{r^2 \left[1 - \frac{v^2}{c^2}\sin^2\psi\right]^{3/2}} \vec{u}_r, \qquad (1)$$

where, $G$ is the universal gravitational constant, $c$ is the speed of light, $\vec{u}_r = \frac{\vec{r}}{r}$, and $\psi$ is the angle between the vectors $\vec{v}$ and $\vec{u}_r$.

He noted that according to this equation, with increasing velocity of the point mass, its gravitational field along the line of sight of the observer to the moving point, *as perceived by the observer*, becomes stronger with the component of this velocity normal to this line ($\psi = \pi/2$) and weaker with the velocity along this line. This effect is just like the electric field of a uniformly moving point charge, and is usually shown by the density of field lines, see for example the book of J. D. Jackson ([14], pp. 553-556) where we find an analogous formula for $\vec{g}_v$ in electromagnetism, but it could well be seen through the length of the field vectors, as is exposed by Jefimenko ([13], pp. 181-184), that makes a detailed analysis of this effect and the field maps that represent both the time-independent gravitational field and the really important dynamic gravitational field map that a single stationary observer would detect as the mass moves past the observer.

The ideas of Sciama, originally published in 1953, seem to be more recognized although his approach and contributions to cosmology were soon abandoned [15]. He focused mainly on the question of the correct interpretation of the physical foundations of GTR [16] and the fundamental role played by Mach's principle on it, having proposed a quantitative implementation of this principle in which the value of $G$ is not anymore an arbitrary constant and somewhat mysteriously connected to the whole universe.

His proposal takes into account Mach's principle and develops on a constructed analogy between classical electromagnetic induction theory and the inertial induction field from which the standard gravitation constitutes the *static component*, and the other component would be a *local acceleration dependent field*. Sciama spent part of his efforts trying to elucidate the key role played by acceleration and arguing that the velocity wouldn't have any perceptible contribution to inertia.

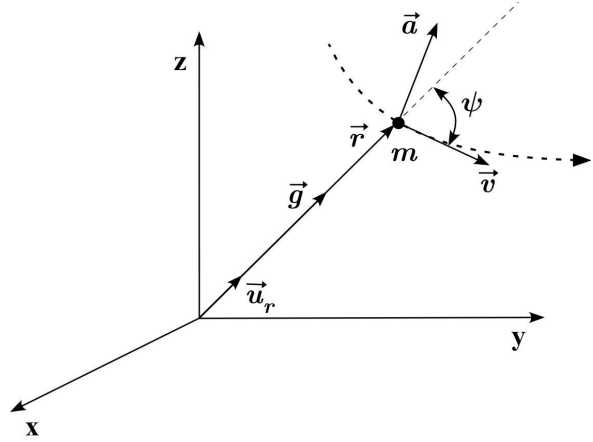

*Figure 1: The general case of the whole inertial induction field $\vec{g}$ of a moving point mass $m$ with velocity $\vec{v}$ and acceleration $\vec{a}$, as seen by the observer at origin.*

As exposed by him, one advantage of Mach's principle over the Newtonian concept of absolute space is that it gives us the opportunity to understand why it is *acceleration* and *not velocity* that can be detected locally, and the inertia itself would be the result of the body acceleration with respect to this local inertial induction field, originated from *distant stars*, which induces the inertial force on the body as if the stars would be truly accelerated opposite to the body. In our point of view this assertion of Sciama is *the announcement of a new symmetry law of nature*.

This change of view supplied by Mach's principle teaches us that *the laws of dynamics may hold in all frames of reference, even non-inertial ones*. The inertial forces that arise in a non-inertial frame would be as a result, physical *local effects* from the distant stars.

Although Sciama didn't propose an explicit dependence of the induction field on the radial component of the acceleration $\vec{a}$ with respect to the observer, it is implicit in his analysis and in the conclusions that he achieved. So we will assume that this is the correct description for the field dependence, and we will discuss this assumption later. Then for our purposes, the field $\vec{g}_a$ induced by an accelerated mass $m$ at some distance $r$ from the observer would take the provisional form:

$$\vec{g}_a = \Phi\, G\, \frac{m}{c^2}\, \frac{\vec{a}_r}{r}, \qquad (2)$$

where $\Phi$ is a coupling constrain parameter and $\vec{a}_r = (\vec{a} \cdot \vec{u}_r)\vec{u}_r$.

Notwithstanding Sciama's considerations against velocity dependence, we have to comment that in our point of view his conclusion is partially correct and will be true only when we consider a *perfectly* homogeneous and isotropic universe as a whole. For each individual interaction the contribution of the relative velocity to the field must be considered. Therefore the total inertial induction field of a moving point mass as experienced by an observer would be the contribution of both components.

To determine $\Phi$, we consider the particular case of radial displacement,

$$\vec{v} = \vec{v}_r = \dot{r}\,\vec{u}_r, \qquad \text{and} \qquad \vec{a}_r = \ddot{r}\,\vec{u}_r. \qquad (3)$$

Thus from equations (1), (2) and (3) the total field in this particular case is:

$$\vec{g} = G\frac{m}{r^2}\left[(1-\frac{\dot{r}^2}{c^2}) + \Phi\frac{\ddot{r}\,r}{c^2}\right]\vec{u}_r. \tag{4}$$

To get the force induced by the field interaction, we settle a test mass $m_g$ at the origin of the reference frame in which will actuate such a resulting force $\vec{F}$, where $k = G\,m\,m_g$, i.e.:

$$\vec{F} = \frac{k}{r^2}\left[1 - \frac{\dot{r}^2 - \Phi\,\ddot{r}\,r}{c^2}\right]\vec{u}_r. \tag{5}$$

If we consider the system as a whole and we impose the conservative condition to it, that the total gravitational energy flux in the system is balanced between kinetic, potential and *radiated*, explicitly: $E_{total} = E_{kinetic} + E_{potential} + E_{radiated}$, this implies necessarily that $\Phi = 2$, since for such a value this interaction can be derived from a *Generalized Potential* $U(r,\dot{r})$ ([17], pp. 227-228) or Schering's potential [18].

To verify this, we take for the generalized potential function, the form:

$$U(r,\dot{r}) = \frac{k}{r}\left(1 + \frac{\dot{r}^2}{c^2}\right). \tag{6}$$

Consequently,

$$\frac{\partial}{\partial r}U(r,\dot{r}) = -\frac{k}{r^2}\left(1 + \frac{\dot{r}^2}{c^2}\right), \tag{7}$$

$$\frac{\partial}{\partial \dot{r}}U(r,\dot{r}) = 2\frac{k}{r}\frac{\dot{r}}{c^2} \longrightarrow \frac{d}{dt}\left(\frac{\partial U}{\partial \dot{r}}\right) = 2\,k\frac{(\ddot{r}\,r - \dot{r}^2)}{c^2\,r^2} \tag{8}$$

By the definition of generalized force:

$$F_r \equiv -\frac{\partial U}{\partial r} + \frac{d}{dt}\left(\frac{\partial U}{\partial \dot{r}}\right). \tag{9}$$

It follows,

$$F_r = \frac{k}{r^2}\left[1 - \frac{\dot{r}^2 - 2\,\ddot{r}\,r}{c^2}\right]. \tag{10}$$

From the expressions (5) and (10), we identify: $\Phi = 2$.

These results show us that it is possible to write a Lagrangian for this interaction field in this particular case. So let's assume for our purposes that the gravitational interaction between material bodies can be described by a field that depends exclusively on their respective distances and radial components of velocities and accelerations, so that we would have a conservative gravitational field derived from a scalar potential, as suggested by Einstein in his lecture [9].

The previously proposed concepts are of fundamental importance in the development of the present work, and the implications that derive from them permit us to establish a formulation that describes the local field related to the *observables*, i.e. the distant points of mass, which manifest their local gravitational influence in terms of their retarded position, velocity and acceleration with respect to us, as we observe them in our own present time, i.e.:

$$\vec{g} = G \frac{m}{r^2} \left[ 1 - \frac{(\vec{v} \cdot \vec{u}_r)^2 - 2\,\vec{a} \cdot \vec{r}}{c^2} \right] \vec{u}_r \,. \tag{11}$$

We can rewrite $\vec{g}$ in terms of the standard static field and a dynamic field:

$$\vec{g} = \underbrace{G \frac{m}{r^2} \vec{u}_r}_{\text{static field}} + \overbrace{G \frac{m}{r^2} \left[ -\frac{(\vec{v} \cdot \vec{u}_r)^2}{c^2} + 2 \frac{\vec{a} \cdot \vec{r}}{c^2} \right] \vec{u}_r}^{\text{dynamic field}} \,. \tag{12}$$

The dynamic field $\vec{g}_D$ has two components, the velocity field and the acceleration field respectively:

$$\vec{g}_v = -G \frac{m}{c^2} \frac{(\vec{v} \cdot \vec{u}_r)^2}{r^2} \vec{u}_r \,, \qquad \text{and} \qquad \vec{g}_a = 2\,G \frac{m}{c^2} \frac{\vec{a} \cdot \vec{u}_r}{r} \vec{u}_r \,. \tag{13}$$

The velocity field $\vec{g}_v$ is essentially a *static field* falling off as $r^{-2}$, in an analogy with electrodynamics ([14], pp. 657-658), whereas the acceleration field $\vec{g}_a$ would be a *gravitational radiation field*, propagating *along* the radius vector and varying as $r^{-1}$.

## 4  The principle of equivalence

In accordance to what was previously established in our framework, let us consider now the general case of a mass immersed in the average local field determined by the whole observable moving masses of a homogeneous and isotropic universe under uniform expansion. Let us apply a force on this test mass $m_g$, say by means of a string, in such a way to produce a net acceleration with respect to a coordinate system *XYZ* fixed on the *distant galaxies* and taken as our stationary reference frame; implicitly the test mass is being accelerated with respect to the average local field. Under these conditions if we assume Mach's principle, a superimposed coordinate system *xyz* (Fig. 2) settled on the mass-string system would be a valid reference frame for the description of the physical situation, and consequently the laws of dynamics must hold.

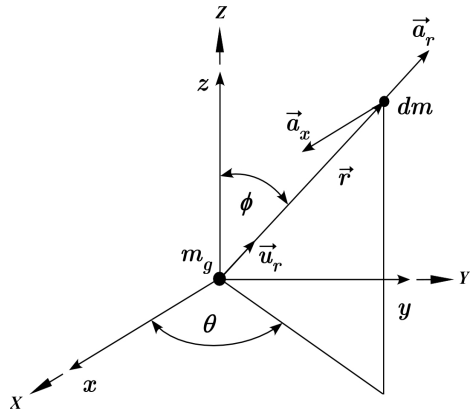

*Figure 2: Accelerations of the mass element $dm$, as seen by the observer at origin of the accelerated reference frame **xyz**.*

From this point of view we observe a resulting acceleration of the whole universe opposite to the test mass, in addition to the radial acceleration of each distant point mass from the uniform expansion.

If we suppose the applied acceleration arbitrarily along the $-X$ direction, the resulting observed acceleration on each mass element $dm$ of the universe would be: $\vec{a}_R = \vec{a}_x + \vec{a}_r$, where $\vec{a}_x = a\,\vec{u}_x$.

The uniform expansion of the universe is expressed mathematically by Hubble's law:

$$\vec{v} = H(t)\,\vec{r}, \tag{14}$$

where $H(t)$ is the Hubble parameter.
This implies that:

$$\vec{a}_r = \frac{d\vec{v}}{dt} = \dot{H}(t)\,\vec{r} + H(t)\,\dot{\vec{r}} = \left(\dot{H} + H^2\right)\vec{r}. \tag{15}$$

From eq. (11) we have that the induced force on the mass $m_g$ by the mass element $dm$ is:

$$d\vec{F} = G\frac{m_g\,dm}{r^2}\left[1 - \frac{\left(\vec{v}\cdot\vec{u}_r\right)^2}{c^2} + 2\frac{\vec{a}_R\cdot\vec{r}}{c^2}\right]\vec{u}_r. \tag{16}$$

When we consider a *perfectly* homogeneous and isotropic universe, we note by the symmetry of the distribution of mass elements that the sum of the whole contributions of them is effective only for the acceleration field, and vanishes for the other field components, in accordance as was argued by Sciama, i.e.:

$$d\vec{F} = 2\,G\frac{m_g}{c^2}\frac{\vec{a}_R\cdot\vec{u}_r}{r}\,dm\,\vec{u}_r. \tag{17}$$

In spherical coordinates the resulting $x$ component of the force element is:

$$dF_x = d\vec{F}\cdot\vec{u}_x = 2\,G\frac{m_g}{c^2 r}\left[a\sin^2\phi\cos^2\theta + \left(\dot{H} + H^2\right)r\sin\phi\cos\theta\right]dm, \tag{18}$$

where $dm = \rho\,r^2\sin\phi\,dr\,d\theta\,d\phi$, and $\rho$ is the average mass density of the universe.

Integrating over all the observable space, i.e. in the limit of the Hubble's radius $R_H$, we get the resulting force on the mass $m_g$ due to the relative acceleration of the whole observable universe:

$$F_x = \iiint_{\substack{over\ all\ the\\ observable\ space}} 2\frac{G}{c^2}m_g\,\rho\,r\sin\phi\left[a\sin^2\phi\cos^2\theta + \left(\dot{H} + H^2\right)r\sin\phi\cos\theta\right]dr\,d\theta\,d\phi, \tag{19}$$

resulting in:

$$F_x = \frac{4\pi G}{3c^2}R_H^2\,\rho\,m_g\,a. \tag{20}$$

Expressing $F_x$ in terms of the critical density $\rho_c$, the density parameter $\Omega$ and the Hubble parameter $H(t)$:

$$F_x = \frac{4\pi G}{3H^2} \rho_c \, \Omega \, m_g \, a \, . \tag{21}$$

By the definition of critical density: $\rho_c \equiv \dfrac{3H^2}{8\pi G}$

We get $F_x$ in its final form:

$$F_x = \frac{1}{2} \Omega \, m_g \, a \, . \tag{22}$$

From the Newtonian point of view, it is the test mass that is being truly accelerated with respect to the reference frame **XYZ**, and reacting with an inertial force $F_I$ proportional to the inertial mass $m_I$ times the applied acceleration $a$, i.e.: $F_I = m_I \, a$, and consequently both forces must be equal:

$$F_I = F_x \, . \tag{23}$$

Then,

$$m_I \, a = \frac{1}{2} \Omega \, m_g \, a \, . \tag{24}$$

Simplifying, we obtain:

$$\frac{m_I}{m_g} = \frac{1}{2} \Omega \, . \tag{25}$$

By the definition of the equivalence principle: $m_I / m_g \equiv 1$, we conclude that: $\quad \Omega = 2$

This is an unexpected result, when we consider the current conviction that $\Omega$ must be $\cong 1$. What does it mean regarding to the present acceleration phase of the universe?

Due in part to the WMAP, which showed the density of matter and energy in the early stages of the universe, most astronomers currently are confident that the universe is flat. But this view is now being questioned by J. Silk and his colleagues at Oxford University, who say it is possible that the WMAP observations have been *interpreted incorrectly*. In a recent article published in Monthly Notices of the Royal Astronomical Society [19], they took data from WMAP and other cosmological experiments and analyzed them using the Bayes' theorem, which can be used to show how the certainty associated with a particular conclusion is affected by different initial assumptions. Using the assumptions of modern cosmology, which assumes *a flat universe* and that *dark energy is a cosmological constant*, they calculated the probability that the universe is in one of three states: flat ($\Omega = 1$), positively curved ($\Omega > 1 \rightarrow$ closed) and negatively curved ($\Omega < 1 \rightarrow$ open). This produced a 98% probability that the universe is indeed flat. However when they performed the calculations again using a more open-minded procedure, i.e. a *curvature scale prior* and a *relaxation of the assumption on the nature of dark energy*, however, the odds changed to 67% in favor of flatness, making a flat universe certainly much less convincing than was previously concluded by the astronomers.

Let us consider now the definition of mass density of the universe:

$$\rho \equiv \frac{3\, M_U}{4\pi\, R_H^3} \ , \tag{26}$$

substituting in the eq. (20) and combining with (23), we find precisely:

$$G = c^2 \frac{R_H}{M_U} \ . \tag{27}$$

Rigorously speaking, this is the scaling law that is a straight consequence of Dirac's large number hypothesis [20], a coincidence that was previously noted by Eddington [21] and hypothesized, albeit not in an explicit form, much earlier by Mach, as commented by Unzicker [22] and Funkhouser [23]. This remarkable relation connecting $G$ with three measurable physical quantities is a necessary condition to be satisfied by any theory that implements Mach's principle, as pointed out by Assis [6].

## 5 Modeling a Galaxy

Our next step is the construction of a dynamic model for a galaxy, without dark matter, and compatible with recent observational data.

To implement our galaxy model (Fig. 3) let us suppose a homogeneous universe in expansion in which there is a region with a higher average density, that constitutes a local attractor centre encircled by its *turnaround* radius $R_t$, in such a way that the matter in its neighborhood is maintained captured by its gravitational field, e.g. a spiral galaxy with a mass density distribution profile highly concentrated on its central bulge, an intermediate region with a decaying density and an external region with a negligible matter density. The application of the previously developed formulation to this simple model, permits us to evaluate the field inside this intermediate region, specifically at any point along a circular orbit of radius $r$, e.g. for fields that are comprised within the range of the galaxy halo. Let us take the galactic centre as the origin of our stationary reference frame *XYZ* fixed on the distant galaxies, and we arbitrarily choose the *X* axis to be coincident with the direction of the vector $\vec{r}$. The dynamic field induced by the mass element $dm$ on the observation point along the *X* axis is:

$$dg_{D_r} = d\vec{g}_D \cdot \vec{u}_r = dg_{v_r} + dg_{a_r} \ . \tag{28}$$

Thus in spherical coordinates we have,

$$\left|\vec{R} - \vec{r}\right| = R\left[1 - 2\left(r\,/\,R\right)\sin\phi\cos\theta + (r\,/\,R)^2\right]^{1/2}, \tag{29}$$

and

$$(\vec{R} - \vec{r}) \cdot \vec{u}_r = R\left(\sin\phi\cos\theta - r\,/\,R\right). \tag{30}$$

Expressing Hubble's law from the centre of mass of the galaxy, which is also the *local dynamic symmetry centre* with respect to the whole expanding universe,

$$\vec{v} = H(t)\,\vec{R} \quad \longrightarrow \quad \vec{a} = \frac{d\vec{v}}{dt} = \left(\dot{H} + H^2\right)\vec{R}\ , \tag{31}$$

Thus from the eq. (13) the dynamic components are:

$$dg_{v_r} = -\frac{G}{c^2}H^2 \frac{\left(1 - (r\,/\,R)\sin\phi\cos\theta\right)^2 \left(\sin\phi\cos\theta - (r\,/\,R)\right)}{\left[1 - 2\left(r\,/\,R\right)\sin\phi\cos\theta + (r\,/\,R)^2\right]^{5/2}} dm\ , \tag{32}$$

and

$$dg_{a_r} = 2\frac{G}{c^2}\left(H^2 + \dot{H}\right)\frac{\left(1-(r/R)\sin\phi\cos\theta\right)\left(\sin\phi\cos\theta-(r/R)\right)}{\left[1-2(r/R)\sin\phi\cos\theta+(r/R)^2\right]^{3/2}}dm, \tag{33}$$

where $dm = \rho\, R^2 \sin\phi\, d\phi\, d\theta\, dR$.

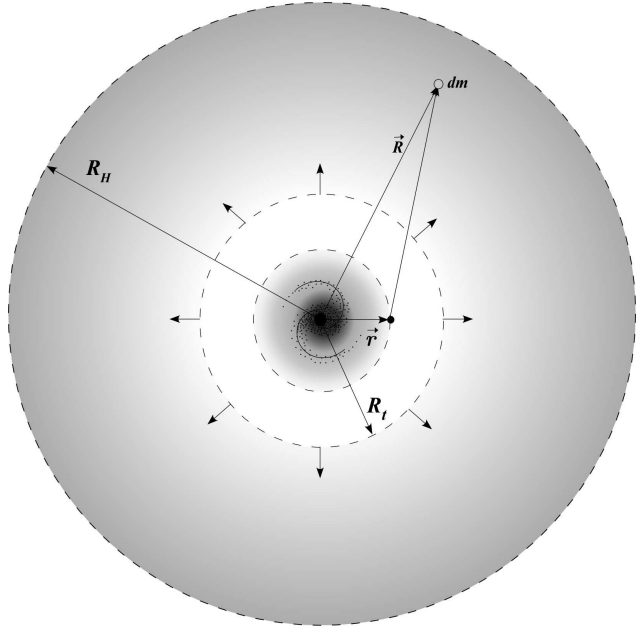

*Figure 3: Galaxy model out of scale: $r \ll R_t \lll R_H$*

Let us impose now the condition that the physically meaningful mass elements are very distant from the observation point, i.e. for $r \ll R$, implicitly we are taking for granted that this assumption is valid for the fields in the range of the galactic halo where we presuppose that the condition $r \ll R_t$ is applicable. Thus, approximating the equations (32) and (33) by a Taylor's series in first order:

$$dg_{v_r} \cong -\frac{G}{c^2}H^2\left(3\frac{r}{R}\sin^2\phi\cos^2\theta + \sin\phi\cos\theta - \frac{r}{R}\right)dm, \tag{34}$$

$$dg_{a_r} \cong 2\frac{G}{c^2}\left(H^2+\dot{H}\right)\left(2\frac{r}{R}\sin^2\phi\cos^2\theta + \sin\phi\cos\theta - \frac{r}{R}\right)dm. \tag{35}$$

For the sake of simplicity in our calculations for the whole *dynamic field*, we will disregard the mass density enclosed between 0 and $R_t$ once their dynamic components, i.e. relative velocities and accelerations, are negligible, and consider only the average mass density of the universe $\rho$ beyond $R_t$ until the limit of Hubble's radius $R_H$. Under these conditions the total dynamic field is estimated by the equation:

$$g_{D_r}(r) \simeq -\frac{G}{c^2} H^2 \rho \int_{R_t}^{R_H} \int_0^{2\pi} \int_0^{\pi} R^2 \sin\phi \left(3\frac{r}{R}\sin^2\phi \cos^2\theta + \sin\phi\cos\theta - \frac{r}{R}\right) d\phi\, d\theta\, dR \; +$$
$$2\frac{G}{c^2}(H^2 + \dot{H}) \rho \int_{R_t}^{R_H} \int_0^{2\pi} \int_0^{\pi} R^2 \sin\phi \left(2\frac{r}{R}\sin^2\phi \cos^2\theta + \sin\phi\cos\theta - \frac{r}{R}\right) d\phi\, d\theta\, dR \,.$$
(36)

The velocity field vanishes in the previous triple integral remaining only the acceleration field component, so that the resulting dynamic field in the interval ($0 \leq r \ll R_t$) is:

$$g_{D_r}(r) \simeq -\frac{4\pi G}{3c^2} \rho\, r\,(H^2 + \dot{H})(R_H^2 - R_t^2)\,. \tag{37}$$

In this equation we observe the negative sign indicating that the resulting field, *in a first instance*, points towards the galaxy centre. This result can be understood by the fact that the system is dynamically asymmetric, i.e. considering the whole observable universe, an observer at any point on the bound orbit around the galaxy centre sees the universe receding radially, not properly from it but from the centre of the galaxy, this slight asymmetry produces a net resultant dynamic field.

Combining this result with the standard static solution we get the total field at a distance $r$ of the galaxy centre:

$$\vec{g}(r) \simeq -G\frac{M_G}{r^2}\vec{u}_r - \frac{4\pi G}{3c^2} \rho\, r\,(H^2 + \dot{H})(R_H^2 - R_t^2)\vec{u}_r\,. \tag{38}$$

Where, by the application of Gauss's theorem to the static field, $M_G$ is the galaxy mass within the sphere of radius $r$.

The previous result can be considered for actual galaxies, where $R_t \ll R_H$. Under this assumption and by the definition of Hubble's radius, it can be expressed in terms of the critical density $\rho_c$ and the density parameter $\Omega$,

$$\vec{g}(r) \simeq -G\frac{M_G}{r^2}\vec{u}_r - \frac{4\pi G}{3H^2}\rho_c\, \Omega\, r\,(H^2 + \dot{H})\vec{u}_r\,. \tag{39}$$

Or, considering again the definition of $\rho_c$ and $\Omega = 2$, we get

$$\vec{g}(r) \simeq -G\frac{M_G}{r^2}\vec{u}_r - r\,(H^2 + \dot{H})\vec{u}_r\,. \tag{40}$$

A test mass $m_g$ positioned at a distance $r$ from the galaxy centre would experience a gravitational pull, toward the centre, of intensity

$$\vec{F}_g(r) \simeq -G\frac{m_g M_G}{r^2}\vec{u}_r - m_g\, r\,(H^2 + \dot{H})\vec{u}_r\,. \tag{41}$$

From the reference frame in the galaxy centre the test mass must spin accordingly to stay in a stable orbit. In the reference frame settled on the test mass it is possible to have two distinctive views, i.e. if we assume the Newtonian point of view, this physical situation is seen as a *virtual* centrifugal force that compensates the attractive gravitational force, and from the point of view of Mach, this physical situation is completely equivalent as if the universe would be revolving in the opposite direction and inducing a *real* centrifugal force. Thus, the forces must be the same and equal to the force of inertia that opposes the gravitational pull. Stated explicitly, $\vec{F}_I = -\vec{F}_g$; this is known as the dynamic equilibrium condition [6].

$$m_I \, a_c \simeq G \frac{m_g M_G}{r^2} + m_g \, r \, (H^2 + \dot{H}). \tag{42}$$

Taking the definition of centrifugal acceleration: $a_c = \omega^2 r$, and rearranging the terms in the equation,

$$m_I \, \omega^2 r - m_g \, r \, (H^2 + \dot{H}) \simeq G \frac{m_g M_G}{r^2}. \tag{43}$$

By the equivalence principle and factoring the expression,

$$\overbrace{m_g \left[ 1 - \left( 1 + \frac{\dot{H}}{H^2} \right) \frac{H^2}{\omega^2} \right]}^{\text{effective inertial mass}} \omega^2 \, r \simeq G \frac{m_g M_G}{r^2}. \tag{44}$$

From the previous result and assuming a dynamic description from the Newtonian point of view, we can define an *effective* inertial mass, $m_{eff}$ for a revolving system with an angular speed $\omega$,

$$m_{eff} \equiv m_g \left[ 1 - \left( 1 + \frac{\dot{H}}{H^2} \right) \frac{H^2}{\omega^2} \right]. \tag{45}$$

That is, for a gravitationally bound system in an expanding universe, a material test body under rotary motion would exhibit a dynamical behavior that could be interpreted as if it would have an effective inertial mass *that reflects the present dynamic state of the universe.*

Recent advances in radio astrometry with the Very Long Baseline Array (VLBA) [24] have shown the potential for precision measurements of the fundamental parameters of the Milky Way galaxy, as the distance to the Galactic centre ($R_0$) and the local speed ($\Theta_0$) of the Local Standard of Rest (LSR). The first estimates indicate a rotation speed of $\Theta_0(R_0) \approx 250$ km s$^{-1}$, some 15% faster than usually assumed and an increase of about 50% in the estimated (dark matter) mass of the Milky Way. The earlier estimate, from simulations, of the content of dark matter inside the sun's orbit was around ~ 60% of the total mass, this would imply from our model an equivalent actual decrease in the effective inertial mass that would generate the same dynamic effect, i.e. $m_{eff}/m_g$ ~ 40%.

With these new observational data, the present estimate of the dark matter content raises to ~ 70% of the totality, and the new corresponding value for the effective inertial mass is $m_{eff}/m_g$ ~ 30%.

Using the standard and not yet so accurate value for $R_0 = 8$ kpc and the up-to-the-minute value of $\Theta_0$, the orbital period of the LSR would be ~ 200 Myr, that implies in an angular speed of $\omega$ ~ $10^{-15}$ rad s$^{-1}$.

Considering the up-to-date value of the Hubble constant $H_0 = 2.30 \times 10^{-18}$ s$^{-1}$ from WMAP, we estimate the ratio: $H_0^2/\omega^2$ ~ $5 \times 10^{-6}$.

From eq. (45) and the previously calculated values, we guess that: $\dot{H}_0 / H_0^2$ ~ $10^5$.

Then eq. (45) can be approximated for the present epoch by:

$$\frac{m_{eff}}{m_g} \simeq 1 - \frac{\dot{H}_0}{\omega^2}. \tag{46}$$

We observe from Fig. 4 that with the increase of $\omega$, the quotient $m_{eff}/m_g$ quickly converges asymptotically to 1 and that the minimum possible angular speed for a revolving star in the edge of the galactic halo to the present epoch is:

$$\omega_{min} \simeq \sqrt{\dot{H}_0} \ . \tag{47}$$

The decrease in the effective inertial mass for $\omega = 5 \times \omega_{min}$ is just 4%, and becomes physically meaningful only in the range: $\omega_{min} < \omega < 5 \times \omega_{min}$. Therefore for massive galaxies, e.g. giant spiral galaxies with high density mass bulge and consequently high internal angular speeds, this effect would become significant only in the external regions. However from the previous analysis we infer that for *low density* dwarf galaxies with very low angular rotation speeds, within the mentioned range, we expect that these systems would be completely subject to this dynamic regime, i.e. they will strongly exhibit the dubbed dark matter *effect* along all its extent.

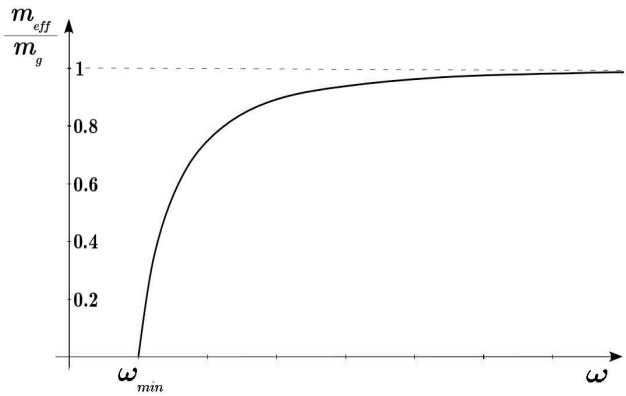

Figure 4: *Effective inertial mass dependence with $\omega$.*

If we consider that the less massive systems rotate with very low angular speeds close to $\omega_{min}$, from the Newtonian point of view this would correspond to a *minimum* inner mass $M_{min}$ inside the revolving radius, consistent with this minimum angular speed. From the dynamic equilibrium condition applied to these systems, this minimum gravitational acceleration would be compensated by a centrifugal acceleration:

$$-\vec{g}_{min} = \vec{a}_c \quad \longrightarrow \quad G\frac{M_{min}}{r^2} = \omega_{min}^2 \, r \, , \tag{48}$$

that is

$$M_{min} = \frac{\omega_{min}^2}{G} r^3 \quad \longrightarrow \quad M_{min} \simeq \frac{\dot{H}_0}{G} r^3 \ . \tag{49}$$

If we consider the radius $r = 300$ pc, we guess $M_{min} \sim 10^7 \, M_\odot$. This estimate matches very well with recent observational data for the Milky Way's dwarf satellite galaxies [25]. The centrifugal acceleration corresponding to this radius for $\omega = \omega_{min}$, is : $a_c \approx 7 \times 10^{-12}$ m s$^{-2}$.

If we think about the Milky Way galaxy considering the plot of $v \times r$ we would not expect to observe a *flat* rotation curve far away from its *visible radius* $R_G$, instead we would see an increasing curve with a very low slope close to $\omega_{min}$, in such way that it exhibits a behavior that emulates a dynamic as if the galaxy would be immersed in a dark matter halo of constant density in the region of the galactic halo, i.e.:

$$\lim_{r \gg R_G} \rho_G \approx \frac{3\dot{H}_0}{4\pi G} \ . \tag{50}$$

These results reproduce, *from first principles*, the phenomenology proposed by M. Milgrom in his theory for the Modified Newtonian Dynamics (MOND) [26], specifically that the Newtonian laws of dynamics (inertia and/or gravity) break down in the limit of small accelerations ($\approx 10^{-10}$ m s$^{-2}$).

The remarkable aspect of these results is that what we misunderstood as a dark matter effect, i.e. a *physically meaningful* decrease in the effective inertial mass, is a direct consequence of the fact that the universe is expanding *accelerating*, specifically whenever $\dot{H}/H^2 \gg 1$.

From this interpretation we are led to conclude that the same energy field that *accelerates* the expansion, explicitly the dark energy, induces in *the gravitationally bound systems* an additional internal field *directed inward*. Concerning this last reasoning the natural conclusion is that one would not observe the dark matter effect in galaxies and clusters in the period immediately preceding the current acceleration phase of the universe, i.e. at a time about 6 billions years before the present epoch. What can one say about the necessity of dark matter in the evolution of the early universe to explain the large-scale structures that we observe today?

## 6  Shedding light on the dark side

As previously commented, in our assumption of the acceleration field dependence on the radial component of the acceleration, we have to point out that this explicit dependence is a *sine qua non* condition to be satisfied by the dynamic gravitational field in such a way to explain the inertia effect of an accelerated material body in what concerns the consistency of the physical foundations of Sciama's ideas, and therefore is a fundamental concept in the present framework of our proposal.

When we take into account a dynamic gravitational field, namely a time varying gravitational field, and its physical effect *along* the vector radius $\vec{r}$ between the observer and the accelerated source, we are truly considering *longitudinal gravitational waves* along this radial direction of propagation, and we must call attention to the understanding that this is a *physical effect* predicted by the GTR as well. However, the introduction of this hypothesis plays a key role in our comprehension of the gravitation, not only pertaining to the explanation of inertia itself but in what also concerns the possibility of our understanding of the dark matter and the dark energy. The effective property that results from this proposition is that, in accordance with Mach's principle, an accelerated material body with respect to an observer induces a dynamic force field that pushes/pulls it in correspondence with the radial direction of the acceleration/deceleration vector of the source relative to the observer. From this remarkable conclusion we are able to understand, for example, that during the acceleration of a material body that moves receding from the observer, it would experience an additional *attractive* dynamic force toward the accelerated body *while the body keeps accelerating*, or oppositely a repulsive dynamic force, if the body decelerates with respect to it. If we apply now these conclusions on an expanding universe centered on us, and think about the receding galaxies *accelerating* from us, we would expect that each one individually is inducing from its radiating field, i.e. the inward longitudinal gravitational waves, an attractive force on us toward it. But if we consider that the universe is homogeneous and isotropic as a whole, then for each galaxy at some distance from us there is an antipode galaxy under the same physical condition with respect to us, i.e. accelerating oppositely and consequently inducing on us an equivalent attractive dynamic force toward it, thus the resulting dynamic field on us vanishes if we are positioned exactly in the centre, and will show a net resulting field if we are slightly shifted from the local attraction centre; this would be the explanation to the present dark matter effect. Let's go further with our analysis and focus now on the inward longitudinal waves radiated by the galaxies, after they have achieved our location in the centre with *no net effect* they will continue now outward, and so after the passing of the required time interval they will reach the corresponding accelerated antipodes galaxies inducing on them now an inward dynamic field, that will promote a slow decrease of the previous acceleration state, and after the elapse of an era, the reversion to a decelerated expansion state that subsequently will generate the next accelerated expansion phase and so forth. This description resembles a stone that falls in the centre of a circular lake with a moving margin, producing concentric waves that move outward and reach the margin pushing/pulling it, depending on the waves' phase, and afterward reverse their movement again toward the centre. About this we note that the longitudinal gravitational waves look like pressure waves in a fluid, which would exert positive/negative pressure along its direction of propagation. Concerning this last reasoning, if we may consider the inflaton field development during the inflationary era, it must have delivered a *finite amount* of energy in the generation process of the longitudinal gravitational waves whose temporal evolution spectra compose the dynamic gravitational field of the universe today, i.e. a Cosmic Gravitational Background (CGB). Where the longitudinal gravitational waves of high frequencies and relative low intensities were probably soon absorbed by the plasma of elementary particles soon after the inflationary period, and this could be a possible explanation for the recent negative result at LIGO (Laser Interferometer Gravitational Wave Observatory) experiment on the detection of gravity waves in these high frequency ranges [27], leaving mostly the stretched ($\lambda \sim 10^{26}$ m) long period longitudinal gravitational waves that continuously drive accelerating/decelerating the universe's expansion in such a way that we would be still living under the influence of the echoes of the Big Bang. Thus if we look back in time we would expect to observe signs of previous accelerated eras. Could we interpret the harmonious patterns of the baryonic acoustic oscillations imprinted in the power spectrum of the Cosmic Microwave Background (CMB) as one of such periods? This interpretation could supply the necessity of dark matter in the early universe to explain the evolution of the large-scale structures to the present epoch, once an accelerated expansion phase necessarily induces an increase in the internal gravitational field of the bound systems, which emulates a dark matter effect intensifying the clump-like formation from the very early density fluctuations. The clumps so formed would stay as strongly bound systems due to the amplification of its internal gravitational field during the accelerated expansion phase. In the subsequent deceleration era after an acceleration phase the formed clumps would suffer now an inertial induction field directed outward due to the deceleration of the expansion ($\dot{H} < 0$), but the clumps would stay still as bound systems due to the fact that the universe continues in expansion and now its size is bigger than in the previous phase; this implies that the energy density carried by the gravitational waves is smaller once they are stretched, an equivalent effect to a red shift; thinking in a quantum mechanical way, the graviton is less energetic and considering too that the size of the clumps remained almost the same size (its cross-section) and that the gravitational wave fronts now have a lower intensity per unit of area due to the expansion. So we would have, as a consequence, an outward lower intensity induced inertial field than in the previous accelerated phase. Additionally, in the acceleration era, the existent plasma would be compressed and consequently through shocks and scattering, converting its kinetic energy into heat. This is an irreversible thermodynamic process and consequently the system would lose part of its energy through the emission of thermal photons, i.e. the observed hot spots in the CMB, leaving at the end the system as a whole in a lower state of gravitational potential energy. Therefore, even if the induced inertial fields were of the same intensity in both eras, in the deceleration phase it would not be strong enough to reverse the formed clumps to the previous dispersed state.

In fact, it would be a coincidence of the present epoch that our estimation for the ratio $H_0^2 / \dot{H}_0 \sim 10^{-5}$ and that the one-in-$10^5$ variations observed in the CMB, is precisely the right amplitude to form the large-scale structures we see today? We don't know!

However, these explanations are consistent with our previous result: that the effective inertial mass of a body under rotary motion reflects the present dynamic state of the universe, i.e. that the present estimate of the dark matter effect is a direct measure of the acceleration rate of the universe's expansion. This reasoning suggests to us that *the main observational evidence of the existence of gravitational waves* is the present accelerated expansion phase of the universe, explicitly *the dark energy field*. Though these considerations are settled on a speculative basis, due yet to the lack of precision observational data, nevertheless they provide a possible known physical origin for dark energy, without resort to any exotic energy field.

As previously mentioned in the introduction, the simplest and immediate explanation for dark energy is Einstein's cosmological constant $\Lambda$, and due to the cosmological observations it must be about $\sim 10^{-52}$ m$^{-2}$ or $(10^{26}$ m$)^{-2}$. This length is not at present related to any other known or expected length scale in nature [28], unless we consider gravitational waves with wavelengths on the order of the radius of the observable universe.

In balance, if we take into account all the previous arguments and consequently we consider the possibility that the WMAP data would be incorrectly interpreted, then the cosmological constant $\Lambda$ would not be actually a *constant*, but rather an oscillating function that decays over the cosmological time due to the continuous dissipation of the energy of the gravitational waves at each acceleration/deceleration cycle, i.e. $\Lambda \to \Lambda(t)$, and the universe could be indeed closed.

This work is not a complete theory of gravity, but rather should be viewed as a potentially interesting proposal that suggests some insights to revise the fundamentals of dynamics and gravitation in such a way to interpret and accommodate the currently observed data.

The interplay of the future theoretical developments and upcoming of high precision observations promises to answer, or at least shed new lights on the key questions still open in this vibrating and ever changing research field.

## Acknowledgments

The author is indebted to many persons who contributed in many ways and forms to the present work, especially Hilda F. Mol, Armando D. Tavares, André K. T. Assis, André L. B. Ribeiro, Camila M. B. Santos, Emico Okuno, Mario A. Tenan, Fermin G. Velasco, Herlon S. Brandão, Patrícia Moreira, Maria C. Lopes, Maria C. do Carmo, Tiago Ventura, Hugo C. B. F. Neves, Angel J. G. Adeva, Dumitru Baleanu, Scott Funkhouser, Diego G. Frias, Alex Y. Ignatiev, Vitor H. da Rosa Bonifácio, Salvatore Capozziello, Grace Vukovich, the reviewers, and all his students over the years.
I wish to express special thanks to Celso F. de Almeida and Agnes M. F. Fausto for encouragement.
I am very grateful to Pedro P. Avelino for the invitation to the postdoctoral stage at Centro de Física do Porto, and the financial support provided by the Universidade Estadual de Santa Cruz during the course of this work.